\documentclass{eas}
\usepackage{graphicx}

\catcode`\@=11
\def\gsim{\ifmmode{\mathrel{\mathpalette\@versim>}}
    \else{$\mathrel{\mathpalette\@versim>}$}\fi}
\def\lsim{\ifmmode{\mathrel{\mathpalette\@versim<}}
    \else{$\mathrel{\mathpalette\@versim<}$}\fi}
\def\@versim#1#2{\lower 2.9truept \vbox{\baselineskip 0pt \lineskip
    0.5truept \ialign{$\m@th#1\hfil##\hfil$\crcr#2\crcr\sim\crcr}}}
\catcode`\@=12

\newcommand{\beq}{\begin{equation}}
\newcommand{\eeq}{\end{equation}}
\newcommand{\az}{{a_0}}
\newcommand{\phiN}{\phi^{\rm N}}
\newcommand{\gv}{{\bf g}}
\newcommand{\gvN}{{\bf g}^{\rm N}}
\newcommand{\xv}{{\bf x}}
\newcommand{\Sv}{{\bf S}}
\begin{document}

\title{N-body simulations in modified Newtonian dynamics} 
\author{Carlo Nipoti}\address{Astronomy Department, University of
  Bologna, via Ranzani 1, I-40127 Bologna, Italy}
\author{Pasquale Londrillo}\address{INAF-Bologna Astronomical
  Observatory, via Ranzani 1, I-40127 Bologna, Italy}
\author{Luca Ciotti$^1$}
\begin{abstract}
  We describe some results obtained with N-MODY, a code for $N$-body
  simulations of collisionless stellar systems in modified Newtonian
  dynamics (MOND). We found that a few fundamental dynamical processes
  are profoundly different in MOND and in Newtonian gravity with dark
  matter.  In particular, violent relaxation, phase mixing and galaxy
  merging take significantly longer in MOND than in Newtonian gravity,
  while dynamical friction is more effective in a MOND system than in
  an equivalent Newtonian system with dark matter.
\end{abstract}
\maketitle
\section{Introduction}

Milgrom~(1983) proposed that the kinematics of galaxies might be
explained without dark matter if one allows for a modification of
Newtonian dynamics in the low-acceleration regime.  In Bekenstein \&
Milgrom's (1984) formulation of Milgrom's modified Newtonian dynamics
(MOND) Poisson's equation $\nabla^2\phiN=4\pi G\rho$ is replaced by
the non-relativistic field equation
 \begin{equation}
\nabla\cdot\left[\mu\left({\Vert\nabla\phi\Vert\over\az}\right)
\nabla\phi\right] = 4\pi G \rho,
\label{eqMOND}
\end{equation} 
with boundary conditions $\nabla\phi\to 0$ as $\Vert\xv\Vert\to\infty$
for a system of finite mass. Here, $\phiN$ and $\phi$ are,
respectively, the Newtonian and MOND gravitational potentials produced
by the density distribution $\rho$, and $\Vert ...\Vert$ is the
standard Euclidean norm. The interpolating function $\mu(y)$ runs
smoothly from $\mu(y)\sim y$ at $y\ll 1$ to $\mu(y)\sim 1$ at $y\gg
1$, with the transition taking place at $y\approx 1$, i.e., when
$\Vert\nabla\phi\Vert$ is of order the characteristic acceleration
$\az \simeq 1.2 \times 10^{-10} {\rm m}\,{\rm s}^{-2}$.  From
Poisson's equation and equation~(\ref{eqMOND}) it follows that the
MOND gravitational field $\gv=-\nabla\phi$ is related to the Newtonian
field $\gvN=-\nabla\phiN$ by ${\mu}(g/\az) \, \gv = \gvN +\Sv$, where
$g\equiv\Vert\gv\Vert$, and $\Sv$ is a solenoidal field dependent on
the specific $\rho$ considered. Unless the system has special
symmetries, $\Sv\ne0$ (Bekenstein \& Milgrom~1984; Brada \&
Milgrom~1995).

Due to the non-linearity of MOND, the problem of calculating the
gravitational field produced by a distribution of $N$ particles is
harder in MOND than in Newtonian dynamics. In MOND there is not an
analytic expression for the force between two particles, so to study
the two-body problem one must solve equation~(\ref{eqMOND}) with
source term given by the two particles (Milgrom~1986). Similarly, when
computing the MOND gravitational potential of a distribution of $N$
particles, one cannot use methods that exploit the linearity of
Poisson's equation. Thus, direct summation or hierarchical (Barnes \&
Hut 1986) Poisson solvers cannot be used in MOND $N$-body codes. As
the solenoidal field $\Sv$ is typically relatively small for
equilibrium configurations (Brada \& Milgrom~1995; Ciotti \etal~2006),
one could be tempted to run $N$-body simulations assuming $\Sv=0$. But
neglecting $\Sv$ when simulating time-dependent dynamical processes
has dramatic effects such as non-conservation of total linear
momentum, as first pointed out by Felten~(1984; see also Nipoti
\etal~2007a).  Instead, equation~(\ref{eqMOND}) must be solved at each
time step.  For collisionless systems a natural choice is to consider
particle-mesh $N$-body codes based on a MOND Poisson solver.

\section{N-MODY: a code for collisionless N-body simulations
in MOND}

We developed N-MODY,\footnote{The code is publicly available upon
  request to the Authors.} a parallel three-dimensional particle-mesh
code for collisionless $N$-body simulations in MOND.  The $N$-body
code and the potential solver on which the code is based are described
in detail in Londrillo \& Nipoti (2008), and have been tested and
applied in Ciotti \etal~(2006, 2007) and Nipoti \etal~(2007a, 2007b,
2007c, 2008). The potential solver of N-MODY solves the MOND field
equation~(\ref{eqMOND}) using a relaxation method in spherical
coordinates based on spherical harmonics expansion. Thus the code is
ideally suited for simulations of isolated stellar systems (but can be
adapted to run simulations of interacting galaxies; see Nipoti
\etal~2007c).  N-MODY is one of the very few MOND $N$-body codes
developed so far: as far as we know the only other three-dimensional
MOND $N$-body code for simulation of collisionless stellar systems is
Brada \& Milgrom's~(1999) code, which is based on a multi-grid
potential solver in Cartesian coordinates and has been implemented
also by Tiret \& Combes~(2007). Recently, a code for MOND cosmological
$N$-body simulations has been developed by Llinares \etal~(2008).

\section{Results of N-body simulations in MOND}

We have applied N-MODY to study a few relevant stellar dynamical
processes, such as collisionless collapse, galaxy merging and
dynamical friction.

Our simulations of collisionless collapses (Nipoti \etal~2007a) showed
that the phase-mixing and violent-relaxation timescales are
significantly longer in MOND than in Newtonian gravity (see also
Ciotti \etal~2007). Remarkably, when MOND systems eventually reach
equilibrium, they have projected surface mass density and velocity
profiles consistent with observations of elliptical galaxies. However,
we found that the collapse end-products cannot satisfy simultaneously
the observed Faber \& Jackson (1976), Kormendy (1977) and Fundamental
Plane (Djorgovski \& Davis~1987, Dressler \etal~1987) relations of
elliptical galaxies, under the assumption of luminosity-independent
stellar mass-to-light ratio.

Our results on phase mixing and violent relaxation in MOND suggested
that galaxy merging could be less effective in MOND than in Newtonian
gravity, also because in MOND galaxies are expected to collide at high
speed, and there are no dark matter halos to absorb orbital energy and
angular momentum (Binney~2004; Sellwood~2004). Our MOND simulations of
collisionless merging confirmed that the merging timescales are
significantly longer in MOND than in Newtonian gravity with dark
matter, suggesting that observational evidence of rapid merging could
be difficult to explain in MOND (Nipoti \etal~2007c).

Finally, we applied N-MODY to study dynamical friction in MOND (Nipoti
\etal~2008).  Our simulations showed that the dynamical friction
timescale is significantly shorter in MOND systems than in Newtonian
systems {\it with the same phase-space distribution of baryons and
additional dark matter}, confirming the analytic estimate of Ciotti \&
Binney~(2004). In Nipoti \etal~(2008) we explored the case of the
evolution a rigid massive bar rotating at the centre of a much more
massive stellar system, but our results are expected to apply also to
the case of a small satellite orbiting within a stellar system, such
as a globular cluster in a galaxy.  As a consequence, in the context
of MOND it is difficult to explain that the globular clusters of the
dwarf spheroidal galaxy Fornax have not sunk yet to the galaxy centre
(see also Ciotti \& Binney~2004 and S\'anchez-Salcedo \etal~2006).

\section{Concluding remarks}

The results of our $N$-body simulations show that MOND systems evolve
dynamically very differently from equivalent Newtonian systems with
dark matter. In particular, phase mixing, violent relaxation and
galaxy merging are less effective, while dynamical friction is more
effective in MOND than in Newtonian gravity with dark matter.

It might seem counterintuitive that in MOND galaxy merging timescales
are long, while dynamical friction timescales are short. However,
galaxy merging cannot be described purely in terms of dynamical
friction. The orbital energy and angular momentum are absorbed mainly
by dark matter halos in the case of Newtonian merging, while by the
outer parts of the baryonic distributions in MOND merging. Thus, in
the Newtonian case there is plenty of particles able to store the
orbital energy of the systems, and the concept of dynamical friction
does apply, while in the MOND case the particles responsible for the
absorption are relatively few, and dynamical friction does not play a
central role (Nipoti \etal~2008).

Similar considerations apply to the case of a bar rotating at the
centre of a MOND galaxy. For the concept of dynamical friction to
apply the bar must rotate in a much more massive stellar system. This
is typically the case in Newtonian systems, because of the presence of
dark matter halos, but not necessarily in MOND: {\it if the bar mass
  is a substantial fraction of the galactic baryonic mass} it might
well be that the slowing-down is slower in MOND than in an equivalent
Newtonian system with dark matter (as actually found by Tiret \&
Combes~2007).

These examples show how a careful comparison between equivalent
systems is fundamental to draw conclusions about dynamical processes
in different gravity theories.  This kind of comparison was the aim of
our $N$-body simulations, which actually showed that MOND systems and
Newtonian systems with dark matter can be distinguished studying their
time evolution. The main result of our experiments is that some
phenomena, such as rapid dissipationless galaxy mergers or the
survival of globular clusters in dwarf spheroidal galaxies, appear
difficult to explain in the context of MOND.


\end{document}